\documentclass[11pt,twoside]{article}
\usepackage{asp2014}

\aspSuppressVolSlug
\resetcounters

\begin{document}

\title{The AFISS web platform for the correlation of high-energy transient events.}

\author{Antonio~Addis$^1$, Andrea~Bulgarelli$^1$, Nicol\`{o}~Parmiggiani$^1$, James~Rodi$^2$, and Angela~Bazzano$^2$}
\affil{$^1$INAF-OAS, Bologna, Italy}
\affil{$^2$INAF-IAPS, Roma, Italy}

\paperauthor{Antonio~Addis}{antonio.addis@inaf.it}{0000-0002-0886-8045}{INAF}{OAS}{Bologna}{Emilia-Romagna}{40129}{Italy}
\paperauthor{Andrea~Bulgarelli}{andrea.bulgarelli@inaf.it}{0000-0001-6347-0649}{INAF}{OAS}{Bologna}{Emilia-Romagna}{4019}{Italy}
\paperauthor{Nicol\`{o}~Parmiggiani}{nicolo.parmiggiani@inaf.it}{0000-0002-4535-5329}{INAF}{OAS}{Bologna}{Emilia-Romagna}{4019}{Italy}
\paperauthor{James~Rodi}{james.rodi@inaf.it}{0000-0003-2126-5908}{INAF}{IAPS}{Roma}{Lazio}{00133}{Italy}
\paperauthor{Angela~Bazzano}{angela.bazzano@inaf.it}{0000-0002-2017-4396}{INAF}{IAPS}{Roma}{Lazio}{00133}{Italy}




  
\begin{abstract}

In the multi-messenger era, facilities share their results with the scientific community through networks such as the General Coordinates Network to study transient phenomena (e.g., Gamma-ray bursts) and implement real-time analysis pipelines to detect transient events, reacting to science alerts received from other observatories. The fast analysis of transient events is crucial for detecting counterparts of gravitational waves and neutrino candidate events. In this context, collecting scientific results from different high-energy satellites observing the same transient event represents a key step in improving the statistical significance of the high-energy candidate events. This project aims to develop a system and a web platform to share information and scientific results of transient events between high-energy satellites with INAF participation (AGILE, FERMI, INTEGRAL and SWIFT). The AFISS platform implements the COMET VOEvent broker  and provides a web portal where the users visualize the list of transient events detected by multi-messenger facilities and received through the GCN. The web portal could show, for each event, a summary of the scientific results shared by the real-time analysis pipelines  and a list of time-correlated transient events. In addition, the  platform is ready to receive results from participating facilities on sub-threshold events (STE) that cannot be shared with the community due to the low statistical significance. If the platform finds a time correlation between two or more STEs, it can promote them to science alerts. The web interface shows the list of STEs with possible time correlation with other STEs or science alerts. The platform notifies the users with an email when a new transient event is received.
  
\end{abstract}

\section{Introduction}

The rapid detection of high-energy transients by multi high-energy satellites represent a key step to increase the significance of the high-energy candidate events, to improve their localization, and to cover the source spectrum over the entire high-energy band. This is crucial for the detection of counterparts of gravitational waves (GWs) and neutrino candidate events.  

The aim of this project is, therefore, to develop a collaborative platform, the AFISS platform,  to enable a joint collaboration between the high-energy (HE) satellites with INAF participation, for prompt e.m. GW and neutrino counterpart searches. This activity is achieved by exchanging alerts from e.m.,  LIGO/Virgo/Kagra collaborations and neutrino facilities, to obtain information of the scientific results generated by the AFISS facilities.

The same collaborative platform is ready to share the results on sub-threshold events between AFISS facilities in the multi-messenger/multi-wavelength context. In the framework of this project some internal tests has been performed with AGILE and INTEGRAL science cases. The next O4 Ligo-Virgo observation period will start in  early 2023 and last for at least one year. A more comprehensive joint use of the multiwavelength information is fundamental to unveil features of high-energy processes, increase the rate of detectable sources by accessing lower amplitude events, improve the estimates of the localization and other important characteristics of the source. This will enable more successful follow-up campaigns of the proposers of the project, and increase the population of sources with identified counterparts. 

The AFISS platform will provide a dashboard through a web portal that will show the list of science alerts received from the GCN network, a summary of the results of the AFISS facilities and a list of sub-threshold events with possible time correlation.

\section{System Design}

\articlefigure{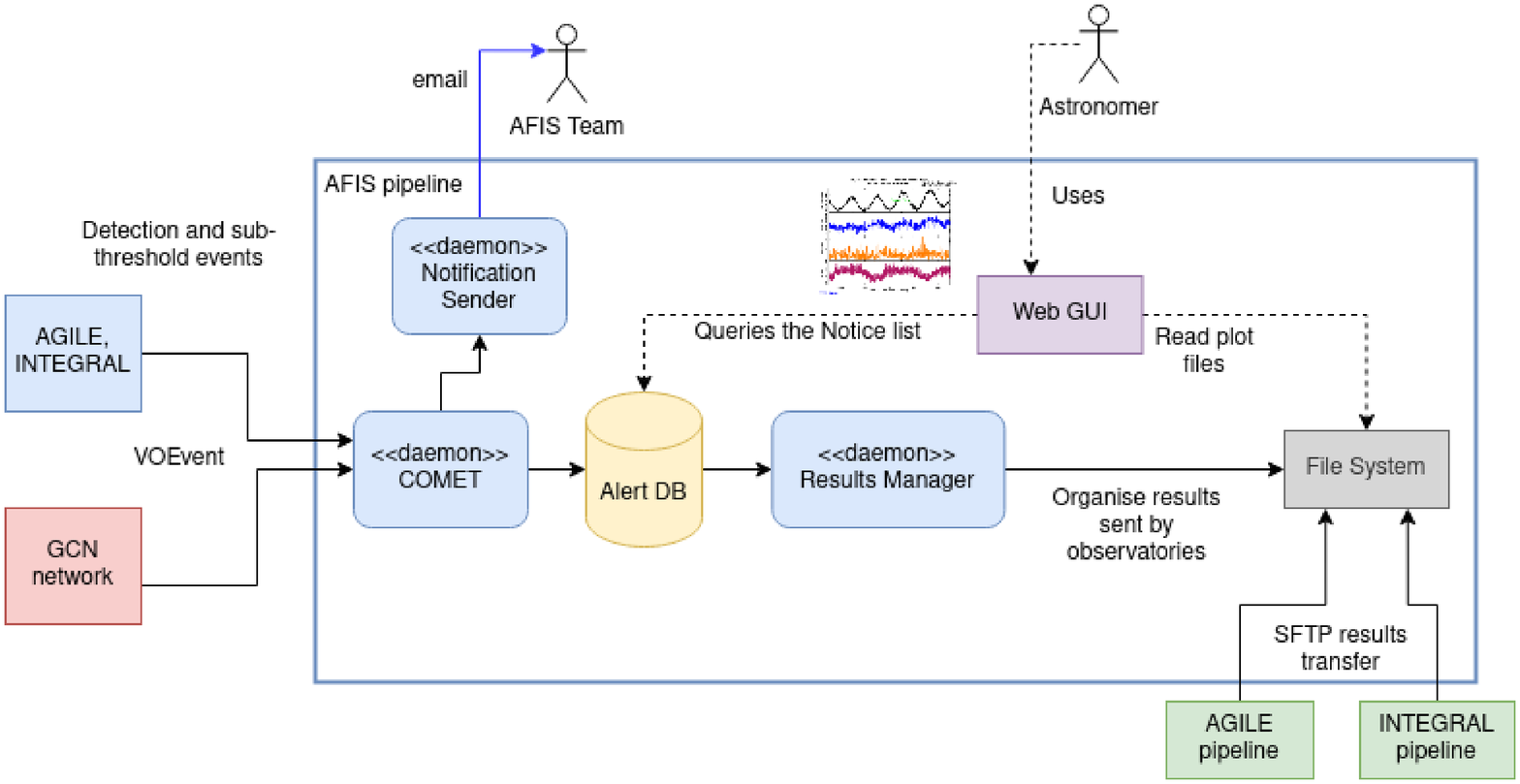}{fig1}{The AFISS Architecture}

The AFISS implements the following components shown in Figure 1:

\begin{itemize}
\item Events Receiver (COMET): this software component is implemented using the COMET framework \citep{https://doi.org/10.48550/arxiv.1409.4805} and aims to receive standard events (Science Alerts) from the GCN network and sub-threshold events from the real-time analysis (RTA) pipelines of the AFISS facilities. These events are then stored in the Alert Database.
\item Notification Sender: this software component sends emails to the AFISS team when a new event is received by the system. It can be configured to notify different types of events.
\item Alert Database: this database stores the standard events and sub-threshold.
\item Results Manager: this software component organizes the scientific results received by the RTA pipelines of the AFISS facilities inside the file system.
\item File System: the file system of the host machine where the AFISS platform is deployed stores the scientific results sent by the AFISS RTA pipelines (e.g., images and plots).
\item Graphical User Interface: the AFISS team can visualize the scientific results and the list of events using this web portal.
\end{itemize}

\section{Data Model}

The input data are VOEvents files: can be science alerts on transient events detected by one of the space mission member of the AFISS project or from external observatories in the Multi-Messenger (MM) / Multi-Wavelength (MW) context (e.g. Gravitational Waves, neutrinos, etc) and shared through different networks.
The platform currently supports AGILE, FERMI\char`_GBM/LAT, INTEGRAL, LVC, MAXI, and ICECUBE notices from GCN Network. In addition it handles sub-threshold events from AGILE\char`_MCAL and INTEGRAL ACS/SPI generated by their pipeline. 
The events are used to promptly identify astrophysical EM events in close temporal coincidence with a standard event or STE.

The AFISS platform generates different types of output:

\begin{itemize}

\item Information extracted from the VOEvents received and stored in the Event Database.
\item Results of the time correlation between STE events that are stored in the Event Database.
\item Emails to send notifications to the AFISS Team

\end{itemize}

\section{Graphical User Interface}

\begin{figure*}[!htb]
	\centering
	  \includegraphics[width=0.7\textwidth]{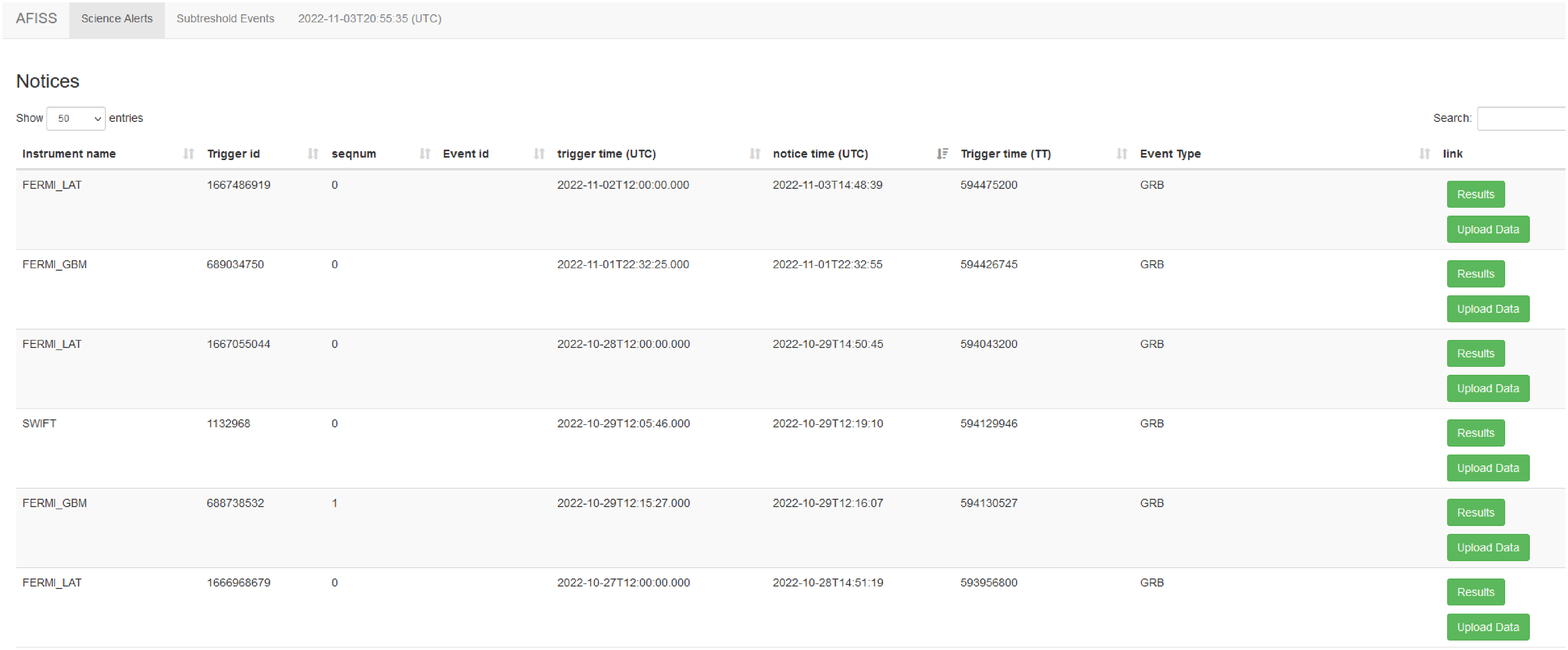}
	  \includegraphics[width=0.2\textwidth]{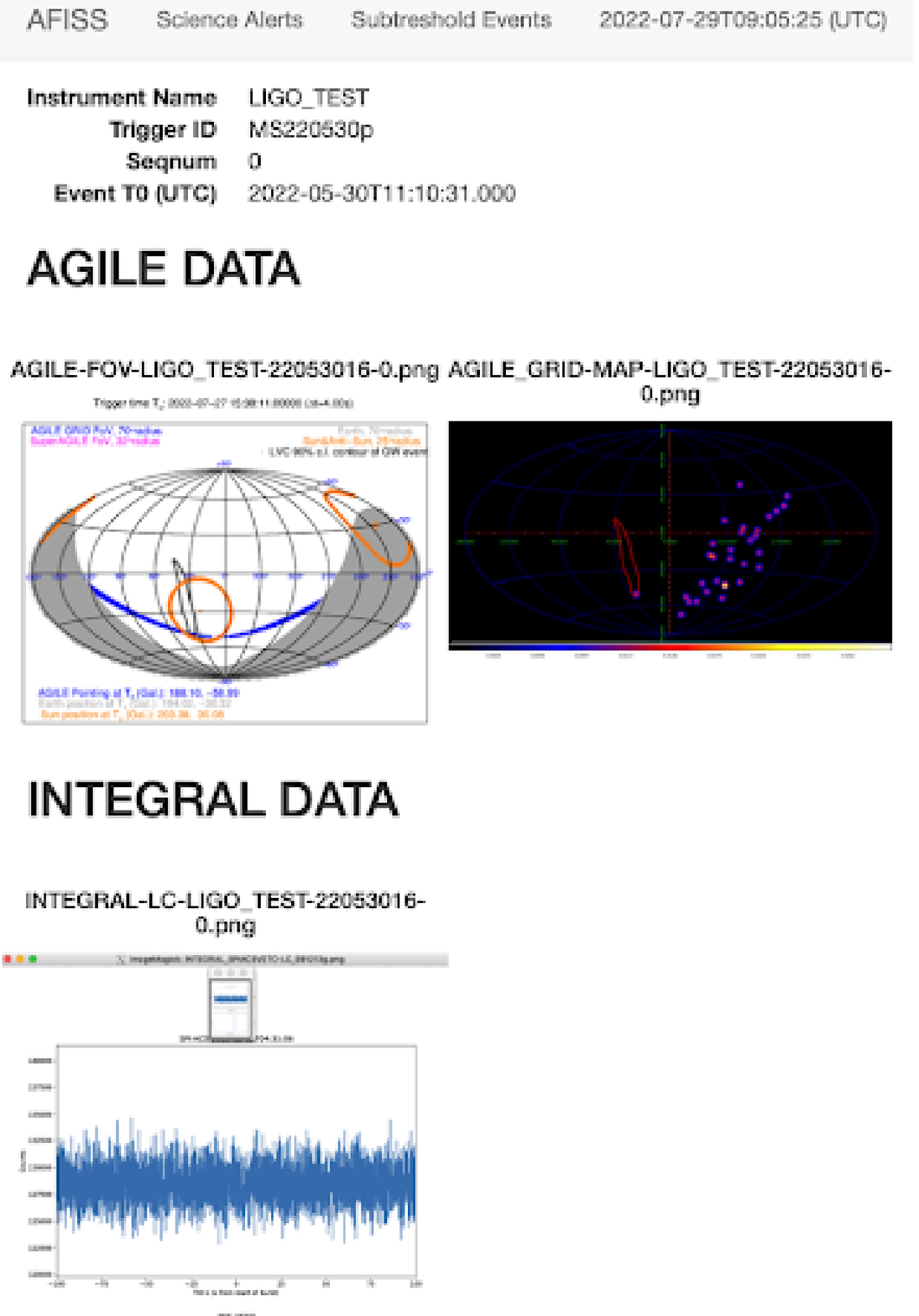}
	\caption{The AFISS home page containing the alerts received from the GCN network, and the data sent by the AFISS team with respect to a specific event}
	\label{fig:architecture}
\end{figure*}
\clearpage
\articlefigure{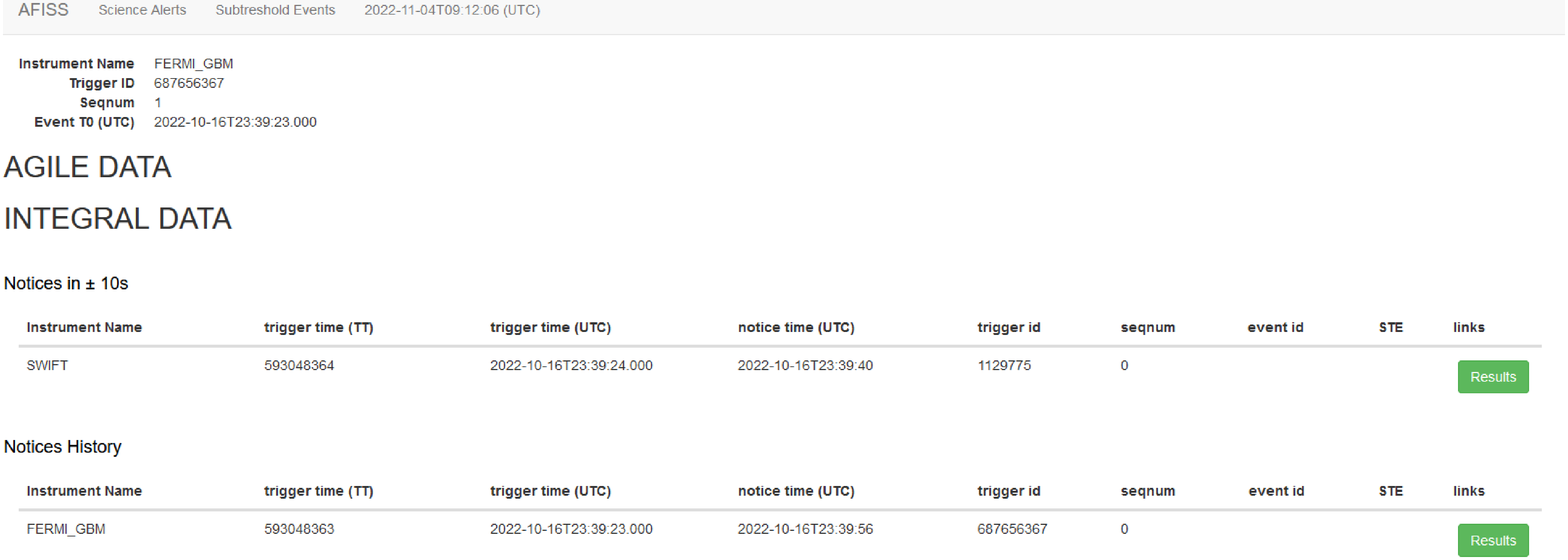}{fig4}{Possible time correlation between SWIFT and FERMI\char`_GBM instruments}

The figures above shows the GUI home page, the data received from the AFISS team and a possible correlation between two instruments. The user can visualize the list of science alerts received from the GCN network and stored in the AFISS database. A specific page is also available to manually update the results from the AFISS facilities.

\section{Conclusions}

The platform is deployed as a Singularity container and it is available to the AFISS team members at http://afiss.iasfbo.inaf.it/afiss/ . New features are under development, such as the addition of new instruments and networks external to the AFISS facilities for public events. The implementation of the new GCN Kafka broker is under investigation.

\section*{Acknowledgments}
The research leading to these results has received funding from the European Union's Horizon 2020 Programme under the AHEAD2020 project (grant agreement n. 871158)


\bibliography{P20}


\end{document}